\documentclass[conference]{IEEEtran}
\IEEEoverridecommandlockouts
\usepackage{cite}
\usepackage{amsmath,amssymb,amsfonts,amssymb}
\usepackage{algorithmic}
\usepackage{graphicx}
\usepackage{textcomp}
\usepackage{parskip}
\usepackage{lettrine}
\usepackage{braket}
\usepackage{xcolor}
\usepackage{subfig}
\usepackage{soul}
\usepackage{comment}
\allowdisplaybreaks 

\begin{document}

\title{Adaptive Measurement-Device-Independent Quantum Key Distribution
\thanks{Mah Noor, Quaid-e-Azam Islamabad, Pakistan. Email: anwar.mahnoor@yahoo.com}
\thanks{A.H. Toor, Quaid-e-Azam University, Islamabad, Pakistan; Email: ahtoor@qau.edu.pk}
}

\author{Mah Noor and A.H. Toor
}
\maketitle
\begin{abstract}
In theory, quantum key distribution (QKD) promises a secure unconditional generation of the key between two remote
participants, based on the laws of quantum physics. However,
 due to limitations in the real-life implementation
of QKD, this promise may fall short of its potential. To bridge the gap between the theory and implementation of QKD, different protocols have
been proposed. One of them is measurement device independent
(MDI) QKD, which can be implemented using present-day
technology and generates a reasonable key rate. This protocol works perfectly fine for intracity communication. However,
the transmission distance is not enough for intercity communication. The other is the adaptive measurement
device-independent (AMDI) QKD that even supports intercity communication.
It holds the promise to replace the present-day classical communication network
such as the Internet, and even quantum communication in the future. In the most recent, implementation of AMDI-QKD, threshold detectors were used.  In this paper, we proposed the replacement of threshold detectors with photon-number resolving detectors. 
We showed that with PNR detectors, the secret key rate ($\mathrm{log_{10} R}$) becomes 11.17 bits/sec, while in the original protocol  it was 10.78 bits/sec. However, this change does not affect the security of the protocol. \\

\end{abstract}

\section{Introduction}

\lettrine[loversize=0.5]{T}oday's economy is based on Internet. The security of Internet dependent on the Rivest-Shamir-Adleman (RSA) algorithm \cite{b1}, which depends on finding prime factors of a very large number, mathematically, a very difficult task with existing computing technologies. However, with quantum computer based Shor's algorithm \cite{b2}, prime factorization of a large number can be achieved relatively easily, hence, undermines the RSA based security of existing communication networks.\\ 
The desired communication security needs to be based on some fundamental law, robust enough to withstand any technological challenges in future. One of such protocols is, the Quantum Key Distribution (QKD), proposed by Charles Bennett and Gilles Brassard in 1984, commonly known as the BB84 protocol \cite{b3}. The security of this protocol is based on no-cloning theorem which makes it impossible to copy an unknown quantum state \cite{b4}.\\
In practical implementation of BB84 protocol, requirements include a single-photon source, zero dark count detectors with unit detection efficiency, infinite dead time and a passive channel. Presently, detector technology falls short of these requirements and hence leads to the possibilities of number of attacks on BB84 setup, such as, large pulse attack \cite{b5}, faked state attack \cite{b6}, time-shift attack \cite{b7}, phase remapping attack \cite{b8}, saturation attack \cite{b9}, photon number splitting attack \cite{b10}, detector blinding attack \cite{b11}, and high power damage attack \cite{b12}. \\
Keeping in mind the limitations of existing detectors, several solutions have been proposed to counter attacks on BB84 setup with a appropriate key transmission rate. Most promising among such solutions is Adaptive Measurement Device Independent Quantum Key Distribution (AMDI-QKD)\cite{b13, b14, b24}. The transmission distance achieved via AMDI-QKD is large enough that it can be used for intercity communication, hence it can replace classical internet with its quantum version.\\
The quantum internet \cite{b23} will also be advantageous because it will be able to teleport
bits between different quantum computers. A qubit can store much
more information than a classical bit, so a huge amount of information
will be able to transmit. Lastly, a quantum computer will
be able to solve complicated computational problems, which even
a supercomputer (to date) cannot do. This is a promised future of
unconditional security and unlimited computing power.\\ 
In the original proposal of AMDI-QKD \cite{b13}, the untrusted relay (Charlie, that could be Eve or a collaborator of her) is assumed to have access to threshold detectors that was restricting his power. In this paper, we replaced the threshold detectors with realistic photon
number resolving detectors (PNR) detectors i.e., we considered
the detectors can count the number of photons in a signal \cite{b15}. This modification increases the supremacy
and power of Charlie (Eve) as single photon sources are used so
he (she) will be able to discriminate between an actual signal sent
by the legitimate users and a signal which got affected due to the
dark counts of detectors as the affected signal will contain of more
than one photon, hence it is a step towards the realistic implementation
of AMDI-QKD \cite{b13}.

\section{Protocol}
The protocol includes the following steps:
\begin{enumerate}
    \item Alice and Bob prepare entangled photons in state $\ket{\psi}$ given as in Eq. \eqref{eq1} and transmit them to Charlie via a quantum channel.
    \begin{equation}\label{eq1}
\mathrm{\ket{\psi} = \frac{1}{\sqrt{1+r^2} }[r\hat{a}_{H}^{\dagger}\hat{c}_{H}^{\dagger} + \hat{a}_{V}^{\dagger}\hat{c}_{V}^{\dagger}
]\ket{0} }.
\end{equation}
Charlie uses photon-number resolving (PNR) detectors, the
POVM for PNR detectors is given as:
\begin{equation}\label{eq2}
\mathrm{\Pi_{k} = \sum_{l=k}^{ \infty}P_{(k|l)}\ket{l}\bra{l} },
\end{equation}
here $\mathrm{P_{(k|l)}}$ i.e., the conditional probability that is given as,
\begin{equation}\label{eq3}
\mathrm{P_{(k|l)}=\sum_{d=0}^{k}
\begin{pmatrix}
  l \\
  k-d
\end{pmatrix} (p_{dark})^d\eta^{k-d}(1-\eta)^{l-k+d},
}
\end{equation}
here, $\eta$ represents the detector's efficiency and $\mathrm{p_{dark}}$ represents the probability of dark counts.
\begin{figure}[h!]
\centering
  {\includegraphics[width=0.5\textwidth]{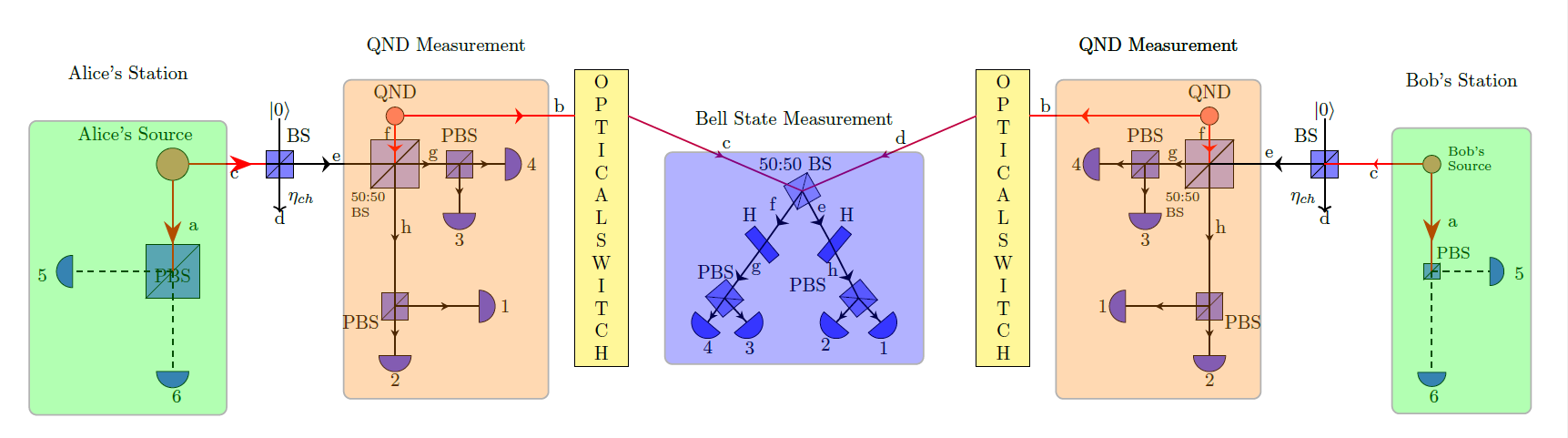}}
  \caption{The schematic layout of AMDI-QKD. The optical devices involved are 50:50 BS, H gate, PBS, and PNR detectors. In QND Measurement, the states $\mathrm{\ket{\psi^{-}}}$ and $\mathrm{\ket{\psi^{+}}}$ can be discriminated. When there is a
successful projection on a Bell state, the input is teleported to the
output mode. In Bell State Measurement, the states $\mathrm{\ket{\psi^{-}}}$ and $\mathrm{\ket{\phi^{-}}}$ can be distinguished.}
  \label{figure1}
\end{figure}
    \item Charlie performs a quantum non-demolition (QND) measurement to confirm the arrival
of photons, i.e., he detects the presence of the photons without
disturbing their states as shown Fig.~\ref{figure1}. The scheme of QND measurement \cite{b16} is based on quantum teleportation
\cite{b17, b18, b22}. The Bell State's Measurement distinguishes between $\ket{\psi^{+}}$ and $\ket{\psi^{-}}$ states. The clicks on detectors 1 and 4 or 3 and 2 correspond
to the projection on state $\ket{\psi^{-}}$, and the clicks on detector 1 and 2 or 3 and 4 corresponds to the projection on state $\ket{\psi^{+}}$. The
clicks on detector 5 or 6 corresponds to the preparation of photon
in H or V component. So there are four events corresponding to the
successful measurement of QND. A successful QND measurement is given as ( $\Pi_{qnd}$):
\begin{equation}\label{eq4}
\Pi_{qnd} = \Pi^{g_{h}}_{1} \otimes \Pi^{g_{v}}_{1} \otimes
\Pi^{h_{h}}_{0} \otimes\Pi^{h_{v}}_{0} \otimes \Pi^{a_{h}}_{1} \otimes \Pi^{a_{v}}_{0}.
\end{equation}
\item Then Charlie pairs the successfully arrived photons via optical
switches  \cite{b13} and performs a Bell State's Measurement (BSM). A successful BSM corresponds to 4 detection events. The BSM provided in Fig. \ref{figure1} distinguishes between $\ket{\psi^{-}}$ and $\ket{\phi^{-}}$ states. The clicks on detetctor 1 and 4 or 3 and 2 correspond to the
projection on state $\ket{\psi^{-}}$ and the clicks on detector 1 and 2 or 3 and
4 correspond to the projection on state $\ket{\phi^{-}}$. Alice or
Bob needs to operate the bit-flip operator on their bits to correlate
them except when they chose Z-bases and Charlie’s BSM corresponds
to the state $\ket{\phi^{-}}$.
\item Charlie then announces via an authentic public channel the outcomes
of his measurement.
\item Alice and Bob keep the data corresponding to the events for
which Charlie gets output and discard the rest.
\item Then Alice and Bob will announce via an authentic public
channel their bases and they select the events where they both
used the same bases.
\item Now to correlate their bases correctly, one of them (Alice or
Bob) needs to operate the bit-flip operator depending upon
the setup and the outcome of Charlie’s measurement.
\item After that Alice and Bob proceed with the usual classical postprocessing
procedure.

\end{enumerate}
\section{Key Rate}
For a realistic setup, the key rate is given as:
\begin{equation}\label{eq5}
\mathrm{R= R_{sifted}[1-H(e_{X})-H(e_{Z} ]}, 
\end{equation}
where $\mathrm{R_{sifted}}$ is the rate of sifted key and $\mathrm{e_X}$ and $\mathrm{e_Z}$ and  corresponds to
bit and phase error rates, respectively \cite{b19}. For this protocol, $\mathrm{R_{sifted}}$ depends on four things:
\begin{enumerate}
\item Channel’s Transmittance $\eta_{ch}$,
\item Probability of success of Bell’s state measurement $p_{BM}$,
\item Probability of success of QND measurement $p_{qnd}$,
\item Source’s Efficiency $\eta_{S}$.
\end{enumerate}
since all these probabilities are independent, so $\mathrm{R_{sifted}}$ will be given
as:
\begin{equation}\label{eq6}
\mathrm{R_{sifted}= \eta_{ch}p_{BM}p_{QND}\eta_{S}.} 
\end{equation}
The total key rate will be:
\begin{equation}\label{eq7}
\mathrm{R= p_{s}p_{BM}[1-H(e_{X})-H(e_{Z} ] ,} 
\end{equation}
where, $\mathrm{p_{s}=\eta_{ch}p_{QND}\eta_{S}}$ is the success probability of both Charlie’s
QND measurement after the photons has arrived through the lossy
channel and Alice’s or Bob’s Z measurement.\\
There are four events corresponding to QND successful measurement
as in figure 1, there are clicks on detectors 1 (horizontal
H component) and 4 (vertical V component), or 3 (H component)
and 2 (V component), or 1 and 2, or 3 and 4 and corresponding
to success measurement just two events, either on click detector 5
(H component) or 6 (V component), so an event corresponds to
the success probability of $p_{s}$ of clicks on detectors 1, 4 and 5. For
each successful event of QND, there are two successful events of Z measurement,
so there are eight different possibilities corresponding
to the success probability of $p_{s}$. All these possibilities are independent
and the probability of success of QND is independent of the
choice of bases by Alice (X or Z), we can write:
\begin{equation}\label{eq8}
\mathrm{ p_{s}=8p_{qnd} .} 
\end{equation}
Bell State Measurement is performed after QND and Z-measurement, let $\mathrm{p_{c}^{Z}}$ be the
probability of the detection events where no one need to operate
the bit flip operator i.e., the projection on state $\ket{\phi^{-}}$ a detection
event corresponds to the clicks on detectors 1 (H component) and 2
(V component) or 3 (H component) and 4 (V component), this is detection pattern after QND, so for Bell's State Measurement alone, it will be $\mathrm{\frac{2p_{c}^{Z}}{p_{qnd}^{2}}}$
. Similarly, $\mathrm{p_{nc}^{Z}}$ is the probability of the detection events where
no one needs to operate the bit flip operator, in case of z-basis it is
just on the state i.e., the projection on state $\ket{\psi^{-}}$ a detection event
corresponds to the clicks on detectors 1 and 4 or 3 and 2, this is
detection pattern after QND, so for Bll's State Measurement alone it will be $\mathrm{\frac{2p_{nc}^{Z}}{p_{qnd}^{2}}}$
. So
$\mathrm{p_{BM}}$ is given as,
\begin{equation}\label{eq9}
\mathrm{ p_{BM}=\frac{2(p_{c}^{Z}+p_{nc}^{Z})}{p_{qnd}^{2}} .} 
\end{equation}
In Z-bases, an error corresponds to the clicks on detectors 1 and 2 or 3 and 4 $\mathrm{p_{nc}^{Z}}$ when Alice and Bob prepare different states or clicks
on detectors 1 and 4 or 3 and 2 $\mathrm{p_{c}^{Z}}$
when they prepare same state.
Similarly, in X-bases it corresponds to events 1 and 2 or 3 and 4 or
1 and 4 or 3 and 2 i.e., $\mathrm{p_{c}^{X}}$ when they prepared photons in the same
bases. So the error rate is given as \cite{b20}:
\begin{equation}\label{eq10}
\mathrm{ e_{Z}=\frac{p_{nc}^{Z} }{p_{c}^{Z}+p_{nc}^{Z} }}, 
\end{equation}
and,
\begin{equation}\label{eq11}
\mathrm{ e_{X}=\frac{p_{nc}^{X} }{p_{c}^{X}+p_{nc}^{X} }}, 
\end{equation}
here $\mathrm{p_{nc}^{X}}$ corresponds to events of successful Bell State Measurement, when Alice and
Bob prepare photons in different states in X-bases. Now the key
rate will be:
\begin{equation}\label{12}
\resizebox{0.4\textwidth}{!}{$
\mathrm{R= \frac{16(p_{c}^{Z}+p_{nc}^{Z})}{p_{qnd}^{2}}[1-H(\frac{p_{nc}^{X} }{p_{c}^{X}+p_{nc}^{X} })-H(\frac{p_{nc}^{Z} }{p_{c}^{Z}+p_{nc}^{Z} }) ] ,} 
$
}
\end{equation}
The probability of a successful QND measurement 
is given as:
\begin{equation}\label{13}
\mathrm{p_{qnd} =  \sum_{z \in \Omega} \tilde{B}_{qnd}(z) \mathit{A}_{(r, \eta_{ch},\eta)}\mathcal{F}^{qnd}},
\end{equation}
\\
where $ \mathit{A}_{(r, \eta_{ch},\eta)}$ represents the channel attenuation factors and $\mathcal{F}^{qnd}$ denotes the detector response functions. The detailed definitions of these two factors along with the combinatorial coefficient $\tilde{B}_{qnd}$ over the summation space $\Omega$ are provided in Appendix A.
\\
Then we performed a Bell State Measurement on the photons. We passed them through a
50:50 beam splitter then polarizing beam splitter, and then measured them. The POVM for the
detection pattern which correspond to the event where neither of
them has to operate the bit-flip operator is given as,
\begin{equation}\label{eq14}
\Pi_{c}^{Z} = \Pi^{g_{h}}_{1} \otimes \Pi^{g_{v}}_{1} \otimes
\Pi^{h_{h}}_{0} \otimes\Pi^{h_{v}}_{0}, 
\end{equation}
here the measurement is done after the operation of optical
switches, before this step Charlie used optical switches, so the detection
efficiency of the detectors will be given as $\eta'=\eta.\eta_{SW}$ where $\eta$ is efficiency of Charlie's detectors and $\eta_{SW}$ is the efficiency of optical switch. The probability of success corresponding to the detection events
after Bell State Measurement where no one needs to operate the bit flip operator in Z-basis is given as,
\begin{equation}\label{eq15}
\begin{split}
p_{c}^{Z}=\underset{z\epsilon\Omega_{z}}{\varSigma}\tilde{B_{c}^{z}}(z).\mathcal{A}(r,\eta_{ch},\eta,\eta').\mathcal{I}_{Z},
\end{split}
\end{equation}
where $\tilde{B}_{c}^{Z}(z)$ is the combinatorial weight for a specific pathway z, $\mathcal{A}_{Z}$ represents the system attenuation and squeezing coefficients, and $\mathcal{I}_{Z}$ is the product of the detector response functions. The explicit definitions of these components and the summation space $\Omega_{Z}$ are detailed in Appendix B. 
\\
The POVM corresponding to a detection event where Alice or Bob
have to operate the bit-flip operator Z-bases is defined as:
\begin{equation}\label{eq16}
\Pi_{nc}^{Z} = \Pi^{g_{h}}_{1} \otimes \Pi^{g_{v}}_{0} \otimes
\Pi^{h_{h}}_{1} \otimes\Pi^{h_{v}}_{0}, 
\end{equation}
and the corresponding probability is:
\begin{equation}\label{eq17}
p_{nc}^{Z}=\underset{z\epsilon\Omega_{nc}}{\varSigma}\tilde{B_{nc}^{z}}(z).\mathcal{G}_{nc}(r,\eta_{CH},\eta,\eta').\mathcal{D}_{nc},
\end{equation}
here, $\tilde{B}_{nc}^{Z}(z)$ represents the combinatorial weighting
of the state, $\mathcal{G}_{nc}$ encapsulates the squeezing and channel
attenuation factors, and $\mathcal{D}_{nc}$ denotes the product of
the detector response functions for the non-coincidence case. The
detailed definitions and summation constraints are provided in Appendix
C.\\

Alice and Bob use X-basis for privacy amplification purposes. The
POVM for the event where neither of them has to operate the bit-flip
operator when using X-basis is given as,
\begin{equation}\label{eq18}
\Pi_{c}^{X} = \Pi^{g_{h}}_{1} \otimes \Pi^{g_{v}}_{1} \otimes
\Pi^{h_{h}}_{0} \otimes\Pi^{h_{v}}_{0}, 
\end{equation}
The probability of success corresponding to detection events where
no one has to operate the bit-flip operator for X-bases is given as:
\begin{equation}\label{eq19}
p_{c}^{x}=\underset{z\epsilon\Omega_{x}}{\varSigma}\tilde{B_{c}^{x}}(z).\mathcal{A}(r,\eta_{ch},\eta,\eta').\mathcal{F}_{x},
\end{equation}
where $\tilde{B_{c}^{X}}(z)$ is the combinatorial weight for a specific
pathway $z$, $\mathcal{A}(r,\eta_{ch},\eta,\eta')$ represents the
system attenuation and squeezing coefficients, and $\mathcal{F}_{X}$
is the product of the detector response functions. The explicit definitions
of these components and the summation space $\Omega_{Z}$ are detailed
in Appendix D.\\
The POVM corresponding to detetction events where no
one has to operate the bit-flip operator for X-bases is given as:
\begin{equation}\label{eq20}
\Pi_{nc}^{X} = \Pi^{g_{h}}_{1} \otimes \Pi^{g_{v}}_{0} \otimes
\Pi^{h_{h}}_{1} \otimes\Pi^{h_{v}}_{0}, 
\end{equation}
and the corresponding probability is given as:
\begin{equation}\label{eq21}
p_{nc}^{x}=\underset{z\epsilon\Omega_{nc,x}}{\varSigma}\tilde{B_{nc}^{x}}(z).\mathcal{G}_{nc}(r,\eta_{CH},\eta,\eta').\mathcal{N}_{nc,x}
\end{equation}
Here, $\tilde{B^{x}}_{nc}(z)$ represents the combinatorial weighting
of the state, $\mathcal{G}_{nc}$ encapsulates the squeezing and channel
attenuation factors, and $\mathcal{N}_{nc,x}$ denotes the product
of the detector response functions for the non-coincidence case. The
detailed definitions and summation constraints are provided in Appendix
E.

Now, the key rate can easily be determined by using Eq.~\eqref{12}
\section{Results}
We studied the dependence of key rate on transmission distance in a logarithmic scale and compared our results with \cite{b20}. The results are shown below:\\
\\
\\
\begin{figure}[h!]
\centering
  \subfloat[Threshold Detectors]{\includegraphics[width=0.35\textwidth]{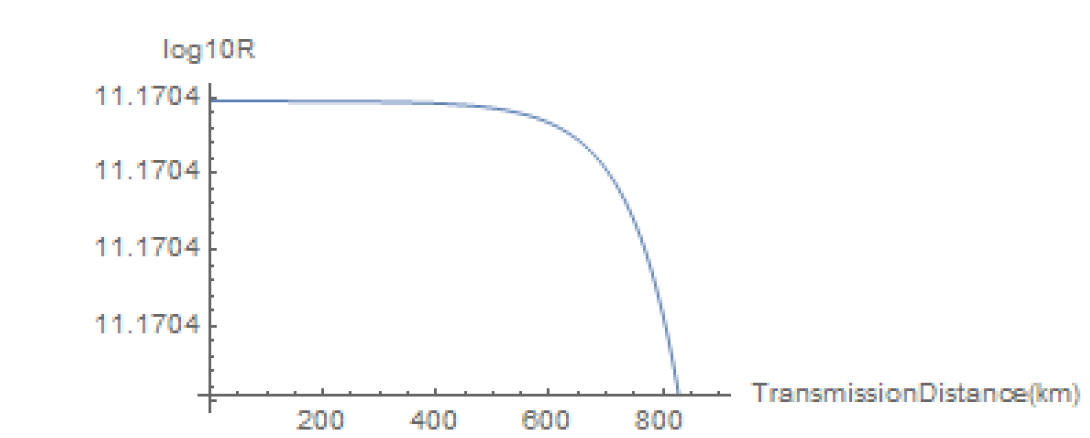}\label{keyrate1}}
  \hfill
  \subfloat[PNR Detectors]{\includegraphics[width=0.35\textwidth]{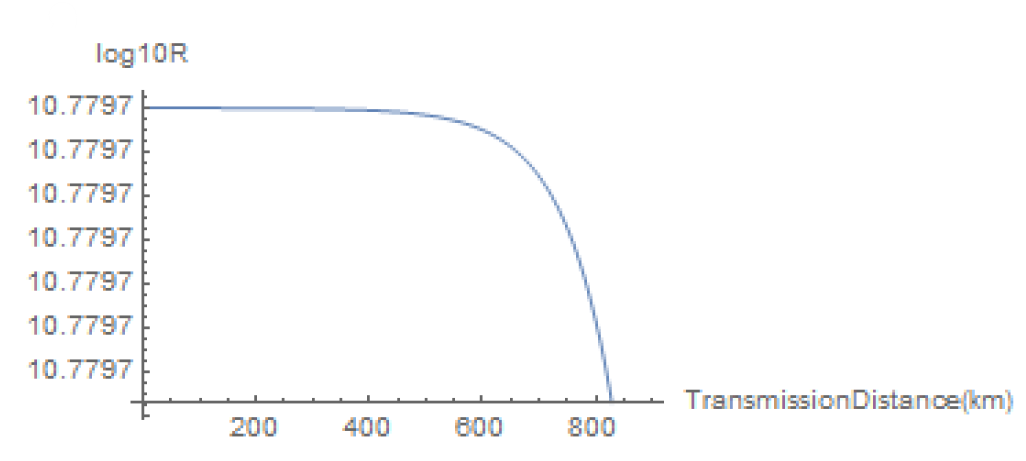}\label{keyrate2}}
  \caption{The secret key rate vs transmission distance AMDI-QKD
when $\mathrm{p_{dark}=2\times10^{-10}}$, $\eta=0.93$, $\mathrm{l_{att}=22\;km}$, $\mathrm{\tau=67\;ns}$ and $\mathrm{c=2\times10^{8}\;ms^{-1}}$. }
\end{figure}\\
As we can see that the replacement of detectors has changed the key rate. The change is not much noticeable. The cut-off distance remained approximately the same.\\
The two vital
parts of key rate are key rate of sifted key and quantum bit error
rate (QBER) which is the ratio of error bits to the total number
of bits transferred. We studied how the replacement of the
detectors affected these two quantities. The behavior of rate of sifted
key with respect to transmission distance is given below.\\
\begin{figure}[h!]
\centering
  \subfloat[Threshold Detectors]{\includegraphics[width=0.35\textwidth]{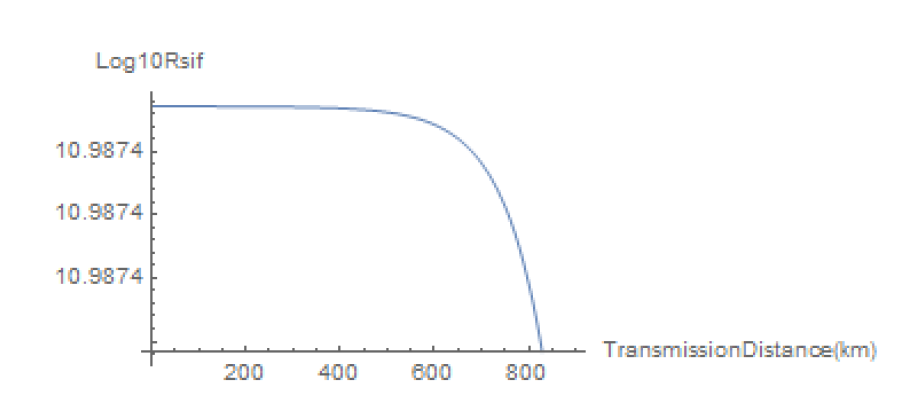}\label{keySrate1}}
  \hfill
  \subfloat[PNR Detectors]{\includegraphics[width=0.35\textwidth]{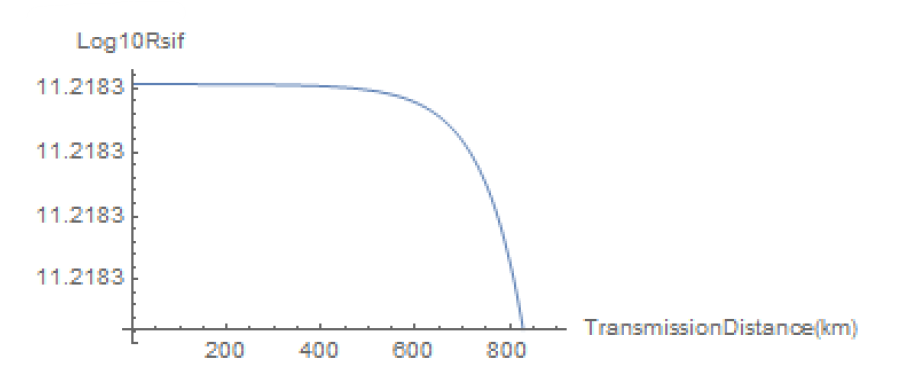}\label{keySrate2}}
  \caption{The sifted key rate vs transmission distance AMDI-QKD
when $\mathrm{p_{dark}=2\times10^{-10}}$, $\eta=0.93$, $\mathrm{l_{att}=22\;km}$, $\mathrm{\tau=67\;ns}$ and $\mathrm{c=2\times10^{8}\;ms^{-1}}$. }
\end{figure}\\
The sifted key rate i.e., the key rate before classical post processing procedures did not
affect much by the replacement of detectors. \\Then we studied
how did QBER  affected by this replacement. The results are shown below.\\
\begin{figure}[h!]
\centering
  \subfloat[Threshold Detectors]{\includegraphics[width=0.3\textwidth]{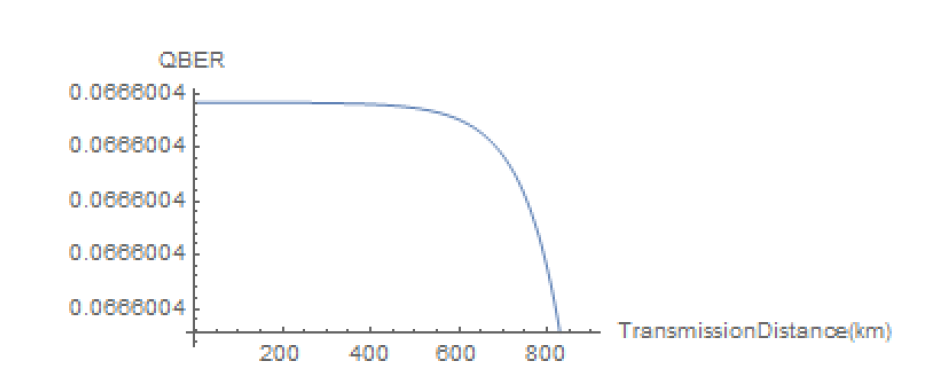}\label{QBER1}}
  \hfill
  \subfloat[PNR Detectors]{\includegraphics[width=0.3\textwidth]{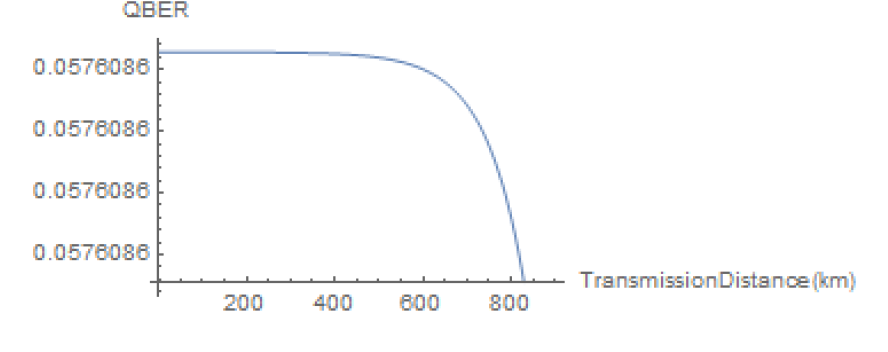}\label{QBER2}}
  \caption{The QBER vs transmission distance AMDI-QKD
when $\mathrm{p_{dark}=2\times10^{-10}}$, $\eta=0.93$, $\mathrm{l_{att}=22\;km}$, $\mathrm{\tau=67\;ns}$ and $\mathrm{c=2\times10^{8}\;ms^{-1}}$. }
\end{figure}\\
As it is clear from Fig.~\ref{QBER2} that the replacement of threshold detectors
with PNR ones has drastically changed the QBER. For the
case of PNR detectors, the ratio of error bits to the numbers of
bits transferred is very large as compared to the threshold ones. The
reason behind that is now Charlie can distinguish between the erroneous
and non-erroneous signals which will ultimately cause more
faulty bits.
\subsection*{\textbf{Conclusion}}
As we have seen by replacing threshold detectors with Photon-Number Resolving Detectors, 
there has been no effect on key rate. The cut-off distance is also same for both detectors. There is some on quantum bit error rate. \\
We have summarized our results in the table given below:
\begin{table}[h!]
\centering
\renewcommand{\arraystretch}{1.5} 
\setlength{\tabcolsep}{12pt}      
\begin{tabular}{|c||l|r|} 
\hline
Parameters & Threshold Detectors & PNR Detectors \\
\hline
Secret Key Rate ($\log_{10} R$) & 11.1704 & 10.7797 \\
\hline
Sifted Key Rate ($\log_{10} R_{sif}$)& 10.9874 & 11.2183 \\
\hline
QBER & 0.0666 & 0.0576\\
\hline
\end{tabular}
\caption{The values for Secret Key Rate, Sifted Key Rate and QBER are given when $\mathrm{p_{dark}=2\times10^{-10}}$, $\eta=0.93$, $\mathrm{l_{att}=22\;km}$, $\mathrm{\tau=67\;ns}$ and $\mathrm{c=2\times10^{8}\;ms^{-1}}$. }
\label{table:1}
\end{table}

We modified AMDI-QKD by replacing threshold detectors with PNR detectors. Our modification made the setup
more realistic i.e., by giving more power to an untrusted relay.  The change does not affect the secret key rate because the Bell state measurement
mentioned in AMDI-QKD protocol (which distinguishes between
$\ket{\psi^{-}}$ and $\ket{\phi^{-}}$ restricts the numbers of photons detected by PNR
detectors because of the Hadamard gates which will result in inconclusive
events when two or more photons are projected to the
beam splitter. Our results align with original work on ADMI-QKD \cite{b20}. The protocol can be implemented with present-day technology,
hence it is a step towards the replacement of classical cryptographic
systems with quantum ones. However, it is worth mentioning that performing QND measurement is a challenging task due the limitations of present-day technology \cite{b25, b26}.\\
The modified protocol produces enough key generation rate
so that it can be used to implement intercity QKD. The cut off distances for both protocols
remains the same. This result implies that even with this modification
AMDI-QKD can still be used for intercity communication. We
also studied the behavior of sifted key rate and QBER with respect
to transmission distance for both detectors and concluded that although
the sifted key rate remains the same for both detectors, the
QBER increases which eventually decreases the secret key rate, but
the effect is not too much.\\

\subsection*{\textbf{Further Work}}
In this paper, we studied AMDI-QKD with PNR detectors.
This setup can be made more realistic by assuming imperfect entangled photon sources i.e., the source which is capable of generating multiple-photon
signals, that protocol would be more realistic implementation of AMDI-QKD and our innovation work will be more useful in that settings. In that case Charlie can implement PNS attack \cite{b10} to get
the full key without introducing any error.
The contemporary entangled sources are PDC type-I source and
PDC type-II source, which often act as probabilistic sources which
mean that they generate photons spontaneously \cite{b21}. So an AMDIQKD
protocol based on entangled sources will not give a sufficient
rate, for that we will need to switch to the decoy-state version of
AMDI-QKD using coherent pulses.

\begin{appendices}
\section{}
In Eq. ~\eqref{13}, the values of $ \mathit{A}_{(r, \eta_{ch},\eta)}$, $\mathcal{F}^{qnd}$ and  $\tilde{B}_{qnd}$ are given as:
\begin{equation*}
\tilde{B}_{qnd}=\underset{k\epsilon\kappa}{\Pi}(-1)^{-w-o-w'-o'}\begin{pmatrix}a_{k}\\
b_{k}
\end{pmatrix},
\end{equation*}
here,
\begin{equation*}
\begin{split}
\kappa = \{ & (1,i), (1,i), (1,u), (1,u), (i,x), (1-i,y), (i,x), (1-i,y), \\
& (y,o), (u,s-w), (1-u,t-o), (x,w'), (y,o'), \\
& (u,s-w'), (x,w), (1-u,t-o') \}.
\end{split}
\end{equation*}
The value of $\mathcal{F}^{qnd}$ is given as:
\begin{equation*}
    \mathcal{F}^{qnd}= f^{qnd}_{x+s+u}f^{qnd}_{1+y-t-u}f^{qnd}_{i},
\end{equation*}
where,
\begin{equation*}
\begin{split}
f^{qnd}_{x} = &x\eta(1-\eta)^{x-1}+p_{d}(1-\eta)^{x}.
\end{split}
\end{equation*}
Similarly, the value of $ \mathit{A}_{(r, \eta_{ch},\eta)}$
is given as:
\begin{equation*}
\begin{split}
\mathit{A}_{(r, \eta_{ch},\eta)} &= \frac{r^{2(i+u)}}{(1+r^{2})^{2}} \sqrt{\eta_{ch}^{2x+2y}(1-\eta_{ch})^{2-2x-2y}} \\
&\quad \times \left( \frac{1}{\sqrt{2}} \right)^{2+2x+2y} (1-\eta)^{s+t+1-i}.
\end{split}
\end{equation*}
The summation space $\Omega$ is defined by the variable $z$ is given as:
\begin{equation*}
z=(i,u,x,y,s,w,w\prime,t,o,o\prime),
\end{equation*}
and,
\begin{equation*}
\Omega=\{z\epsilon Z_{\geq0}^{10}:\mathrm{constraints(C1)-(C10)}\},
\end{equation*}
where the constraints are given as:
\begin{itemize}
\item (C1) $0\leq i\leq1$ 
\item (C2) $0\leq u\leq1$ 
\item (C3) $0\leq x\leq i$ 
\item (C4) $0\leq y\leq1-i$ 
\item (C5) $0\leq s\leq x+u$ 
\item (C6) $max\{0,s-u\}\leq w\leq min\{s,x\}$ 
\item (C7) $max\{0,s-u\}\leq w'\leq min\{s,x\}$ 
\item (C8) $0\leq t\leq1-u+y$ 
\item (C9) $max\{0,t-(1-u)\}\leq o\leq min\{t,y\}$ 
\item (C10) $max\{0,t-(1-u)\}\leq o'\leq min\{t,y\}$ 
\end{itemize}
\section{}
In Eq.~\eqref{eq15}, the detector response function $\mathcal{I}_{Z}$ is given as:
\[
\mathcal{I}_{Z}=f_{x+u-s}^{c,z}f_{X+U-S}^{c,z}f_{1+y-t-u}^{c,z}f_{i}^{c,z}f_{I}^{c,z}f_{1+Y-T-U}^{c,z}f'{}_{a}^{c,z}f'{}_{d-a}^{c,z}
\]

where,

\[
f_{x}^{c,z}=x\eta(1-\eta)^{x-1}+p_{d}(1-\eta)^{x}
\]

and,
\[
f'{}_{x}^{c,z}=x\eta'(1-\eta')^{x-1}+p_{d}(1-\eta')^{x}
\]
The combinatorial factor $\tilde{B}_{c}^{Z}(z)$ is given as:
\begin{equation*}
    \begin{split}
        \tilde{B_{c}^{z}}=&\underset{k\epsilon\kappa_{z}}{\Pi}(-1)^{-w-o-w'-o'-W-O-W'-O'+l'+q'}\\
        &\times(-1)^{j+l+q+J+j'+J'}\begin{pmatrix}a_{k}\\
b_{k}
\end{pmatrix}
    \end{split}
\end{equation*}

here,
{\small
\begin{equation*}
\begin{split}
\kappa_z & = \{  (1,i),(1,i),(1,u),(1,u),(i,x),(1-i,y),(i,x),(1-i,y),\\
& (x,w),(y,o), (u,s-w),(1-u,t-o),(x,w\prime),(y,o\prime),(1,U),\\
&(u,s-w\prime),(1-u,t-o\prime)
(1-u,q),(u,l),(1,I),(1,I),(I,X),\\
&(1-I,Y),(1-I,Y),(X,W),(Y,O),(U,S-W),(X,W'),\\
&(Y,O'),(1-U,T-O),(U,S-W'),(1-U,T-O'),\\
&(U,L),(1-u,d-l-L-q),(U,L'),(u,l'),(l+L,j),(u,l),\\
&(1-U,d-l'-L'-q'),(d-l-L,a-j),(u-l+U-L,J),\\
&(2-u-d+l+L-U,A-J),(u-l'+U-L',J'),\\
&(2-u-d+l'+L'-U,A-J'),
(d-l'-L',a-j')
(1,U),\\
& (1-u,q'),(I,X),(l'+L',j'),(1-u,q)
\}.
\end{split}
\end{equation*}
}
The physical parameters and phase factors, i.e., $\mathcal{A}(r,\eta_{ch},\eta,\eta')$ is given as:
\begin{align*}
\mathcal{A}(r,\eta_{ch},\eta,\eta')=\frac{r^{2(i+u+I+U)}}{(1+r^{2})^{4}}\sqrt{\eta_{CH}}^{2(x+y+X+Y)}(1-\eta')^{2-d}\\
\sqrt{1-\eta_{CH}}^{4-2(x+y+X+Y)}
\frac{1}{\sqrt{12}}^{12+2x+2y+2X+2Y}\\
(1-\eta)^{2+s+t-i+S+T-I}
\end{align*}
The summation space $\Omega_z$ is 33-dimensional, defined by the variable z given as:
\begin{equation*}
    \begin{split}
        z=&(i,u,x,y,s,w,w\prime,t,I,U,X,Y,S,W,W',o,T,O,d,O',\\
        &a,l',o\prime,L',A,l,L,j,q,j',J',q',J)
    \end{split}
\end{equation*}

and,
\[
\Omega_{z}=\{z\epsilon Z_{\geq0}^{33}:\mathrm{constraints(C1)-(C33)}\}
\]
The constraints are given as:
\begin{itemize}
\item (C1) $0\leq i\leq1$ 
\item (C2) $0\leq u\leq1$ 
\item (C3) $0\leq x\leq i$ 
\item (C4) $0\leq y\leq1-i$ 
\item (C5) $0\leq s\leq x+u$ 
\item (C6) $max\{0,s-u\}\leq w\leq min\{s,x\}$ 
\item (C7) $max\{0,s-u\}\leq w'\leq min\{s,x\}$ 
\item (C8) $0\leq t\leq1-u+y$ 
\item (C9) $0\leq I\leq1$ 
\item (C10) $0\leq U\leq1$ 
\item (C11) $0\leq X\leq I$ 
\item (C12) $0\leq Y\leq1-I$
\item (C13) $0\leq S\leq X+U$
\item (C14) $max\{0,S-U\}\leq W\leq min\{S,X\}$ 
\item (C15) $max\{0,S-U\}\leq W'\leq min\{S,X\}$ 
\item (C16) $max\{0,t-(1-u)\}\leq o\leq min\{t,y\}$ 
\item (C17) $0\leq T\leq1-U+Y$ 
\item (C18) $max\{0,T-(1-U)\}\leq O\leq min\{T,Y\}$ 
\item (C19) $0\leq d\leq2$
\item (C20) $max\{0,T-(1-U)\}\leq O'\leq min\{T,Y\}$ 
\item (C21) $0\leq a\leq d$
\item (C22) $max\{0,d-2+u\}\leq l'\leq min\{d,u\}$ 
\item (C23) $max\{0,t-(1-u)\}\leq o'\leq min\{t,y\}$ 
\item (C24) $max\{0,d-2-l'+u+U\}\leq L'\leq min\{d-l',U\}$ 
\item (C25) $0\leq A\leq2-d$
\item (C26) $max\{0,d-2+u\}\leq l\leq min\{d,u\}$ 
\item (C27) $max\{0,d-2-l+u+U\}\leq L\leq min\{d-l,U\}$ 
\item (C28) $max\{0,a-d+l+L\}\leq j\leq min\{a,l+L\}$
\item (C29) $max\{0,d-l-L-1+U\}\leq q\leq min\{d-l-L,1-u\}$
\item (C30) $max\{0,a-d+l'+L'\}\leq j'\leq min\{a,l'+L'\}$
\item (C31) $max\{0,2-u-d+l'+L'-U\}\leq J'\leq min\{A,u-l'+U-L'\}$ 
\item (C32) $max\{0,d-l'-L'-1+U\}\leq q'\leq min\{d-l-L',1-u\}$
\item (C33) $max\{0,2-u-d+l'+L'-U\}\leq J\leq min\{A,u-l'+U-L'\}$ 
\end{itemize}
\section{}
In Eq. \eqref{eq17}, the combinatorial factor 
$\tilde{B}_{nc}^{Z}(z)$ is given as:
\[
\tilde{B_{nc}^{z}}=\underset{k\epsilon\kappa_{nc}}{\Pi}(-1)^{-w-o-w'-o'-W-O-W'-O'+l'+q'+j+l+q+J+j'+J'}\begin{pmatrix}a_{k}\\
b_{k}
\end{pmatrix}
\]
here, 
\small{
\begin{equation*}
    \begin{split}
        \kappa_{nc}=\{&(1,i),(1,i),(1,u),(1,u),(i,x),(1-i,y),(i,x),(1-i,y),\\
&\times (x,w),(y,o),(u,s-w),(1-u,t-o),(x,w\prime),(y,o\prime),\\
&\times (1-u,t-o\prime),(1,I),(1,I),(1,U),(I,X),(1-I,Y),(I,X),\\
&\times (1-I,Y),(u,s-w\prime),(X,W),(Y,O),(U,S-W),\\
&\times (1-U,T-O),(X,W'),(U,S-W'),(1-U,T-O'),\\
&\times(1,U),(u,l),(1-u,q),(1-U,d-l-L-q),(U,L),\\
&\times (Y,O'),(U,L'),(u,l'),(1-U,d-l'-L'-q'),(1-u,q'),\\
&\times (l+L,j),(u-l+U-L,J),(d-l-L,a-j),\\
&\times (d-l'-L',a-j'),(2-u-d'+l'+L'-U,A'-J'),\\
&\times (2-u-d+l+L-U,A-J),(l'+L',j'),\\
&\times(u-l'+U-L',J')
    \end{split}
\end{equation*}}

The physical system parameters and the complex phase interference $\mathcal{G}_{nc}$ is given as:
\begin{equation*}
    \begin{split}
        \mathcal{G}_{nc}(r,\eta_{CH},\eta,\eta')= &\frac{r^{2(i+u+I+U)}}{(1+r^{2})^{4}}\sqrt{\eta_{CH}}^{2(x+y+X+Y)}\\
& \times \sqrt{1-\eta_{CH}}^{4-2(x+y+X+Y)}
\\
& \times \frac{1}{\sqrt{12}}^{12+2x+2y+2X+2Y}(1-\eta')^{2-a-A}\\
&\times (1-\eta)^{2+s+t-i+S+T-I}
    \end{split}
\end{equation*}

The detection response function $\mathcal{D}_{nc}$ is given as:
\small{\[
\mathcal{D}_{nc}=f_{x+u-s}^{nc,z}f_{X+U-S}^{nc,z}f_{1+y-t-u}^{nc,z}f_{1+Y-T-U}^{nc,z}f_{i}^{nc,z}f_{I}^{nc,z}f_{a}^{nc,z}f_{A}^{nc,z}
\]}

where,

\[
f_{x}^{nc,z}=x\eta(1-\eta)^{x-1}+p_{d}(1-\eta)^{x}
\]

and,
\[
f'{}_{x}^{nc,z}=x\eta'(1-\eta')^{x-1}+p_{d}(1-\eta')^{x}
\]

The variable $z$ that defines the summation space is given as:
\begin{equation*}
\begin{split}
    z=&(i,u,x,y,s,w,w\prime,t,o,o',I,U,X,Y,S,W,W',T,d,d',\\
    &O,O',l,a,a',A,A',l',L',L,q'q,j,j',J,J')
\end{split}
\end{equation*}

\[
\Omega_{z}^{nc}=\{z\epsilon Z_{\geq0}^{36}:\mathrm{constraints(C1)-(C36)}\}
\]

\begin{itemize}
\item (C1) $0\leq i\leq1$
\item (C2) $0\leq u\leq1$
\item (C3) $0\leq x\leq i$
\item (C4) $0\leq y\leq1-i$
\item (C5) $0\leq s\leq x+u$
\item (C6) $max\{0,s-u\}\leq w\leq min\{s,x\}$
\item (C7) $max\{0,s-u\}\leq w'\leq min\{s,x\}$
\item (C8) $0\leq t\leq1-u+y$
\item (C9) $max\{0,t-(1-u)\}\leq o\leq min\{t,y\}$
\item (C10) $max\{0,t-(1-u)\}\leq o'\leq min\{t,y\}$
\item (C11) $0\leq I\leq1$
\item (C12) $0\leq U\leq1$
\item (C13) $0\leq X\leq I$
\item (C14) $0\leq Y\leq1-I$
\item (C15) $0\leq S\leq X+U$
\item (C16) $max\{0,S-U\}\leq W\leq min\{S,X\}$
\item (C17) $max\{0,S-U\}\leq W'\leq min\{S,X\}$
\item (C18) $0\leq T\leq1-U+Y$
\item (C19) $0\leq d\leq2$
\item (C20) $0\leq d'\leq2$
\item (C21) $max\{0,T-(1-U)\}\leq O\leq min\{T,Y\}$
\item (C22) $max\{0,T-(1-U)\}\leq O'\leq min\{T,Y\}$
\item (C23) $max\{0,d-2+u\}\leq l\leq min\{d,u\}$
\item (C24) $0\leq a\leq d$
\item (C25) $0\leq a'\leq d'$
\item (C26) $0\leq A\leq2-d$
\item (C27) $0\leq A'\leq2-d$'
\item (C28) $max\{0,d'-2+u\}\leq l'\leq min\{d',u\}$
\item (C29) $max\{0,d'-2-l'+u+U\}\leq L'\leq min\{d'-l',U\}$
\item (C30) $max\{0,d-2-l+u+U\}\leq L\leq min\{d-l,U\}$
\item (C31) $max\{0,d'-l'-L'-1+U\}\leq q'\leq min\{d'-l'-L',1-u\}$
\item (C32) $max\{0,d-l-L-1+U\}\leq q\leq min\{d-l-L,1-u\}$
\item (C33) $max\{0,a-d+l+L\}\leq j\leq min\{a,l+L\}$
\item (C34) $max\{0,a'-d+l'+L'\}\leq j'\leq min\{a',l'+L'\}$
\item (C35) $max\{0,2-u-d'+l'+L'-U\}\leq J'\leq min\{A',u-l'+U-L'\}$
\item (C36) $max\{0,2-u-d+l+L-U\}\leq J\leq min\{A,u-l+U-L\}$
\end{itemize}
\section{}
In Eq. \eqref{eq19} the combinatorial factor $\tilde{B_{c}^{X}}(z)$ is given as:
\[
\tilde{B_{c}^{x}}=\underset{k\epsilon\kappa_{x}}{\Pi}(-1)^{-w-o-w'-o'-W-O-W'-O'+l'+q'+j+l+q+J+j'+J'}\begin{pmatrix}a_{k}\\
b_{k}
\end{pmatrix}
\]

here,
\begin{equation*}
    \begin{split}
        \kappa_{x}=\{&(1,i),(1,i),(1,u),(1,u),(i,x),(1-i,y),(i,x),(1-i,y),
        (x,w),\\
        &\times(y,o),
     (u,s-w),(1-u,t-o),(x,w\prime),(y,o\prime),(1,U),(u,s-w\prime),\\
     &\times (1-u,t-o\prime)
,(1-u,q),(u,l),(1,I),(1,I),(1-I,Y),(I,X),\\
&\times (1-I,Y),(X,W),(Y,O)
(U,S-W),(1-U,T-O),(X,W'),\\
&\times(Y,O'),(U,S-W'),(1-U,T-O'),
(1-u,q),(U,L),(U,L'),\\
& \times(u,l'),(1-u,d-l-L-q),(1-U,d-l'-L'-q')
,(l+L,j),\\
&\times (d-l-L,a-j),(u,l),(u-l+U-L,J),(d-l'-L',a-j'),\\
&\times(2-u-d+l+L-U,A-J),(u-l'+U-L',J'),(1,U),\\
&\times(I,X),(2-u-d+l'+L'-U,A-J'),(l'+L',j'),(1-u,q')
    \end{split}
\end{equation*}

The detection function $\mathcal{F}_{X}$ is given as

\[
\mathcal{F}_{X}=f_{x+u-s}^{c,x}f_{X+U-S}^{c,x}f_{1+y-t-u}^{c,x}f_{i}^{c,x}f_{I}^{c,x}f_{1+Y-T-U}^{c,x}f'{}_{a}^{c,x}f'{}_{d-a}^{c,x}
\]

where,

\[
f_{i}^{c,x}=i\eta(1-\eta)^{i-1}+p_{d}(1-\eta)^{i}
\]

and,
\[
f'{}_{i}^{c,x}=i\eta'(1-\eta')^{i-1}+p_{d}(1-\eta')^{i}
\]
Likewise, the value of $\mathcal{A}(r,\eta_{ch},\eta,\eta')$ encompassing physical parameters is given as:

\begin{align*}
\mathcal{A}(r,\eta_{ch},\eta,\eta')=\frac{r^{2(i+u+I+U)}}{(1+r^{2})^{4}}\sqrt{\eta_{CH}}^{2(x+y+X+Y)}\sqrt{1-\eta_{CH}}^{4-2(x+y+X+Y)}\\
\frac{1}{\sqrt{12}}^{12+2x+2y+2X+2Y}(1-\eta')^{2-d}(1-\eta)^{2+s+t-i+S+T-I}
\end{align*}
The summation space $\Omega_{z}$ is described by the variable $z$ given as
\begin{equation*}
    \begin{split}
        z=&(i,u,x,y,s,w,w\prime,t,I,U,X,Y,S,W,W',o,T,O,d,O',a,l',o\prime,L',A,\\
        &l,L,j,q,j',J',q',J)
    \end{split}
\end{equation*}

and,

\[
\Omega_{z}=\{z\epsilon Z_{\geq0}^{33}:\mathrm{constraints(C1)-(C33)}\}
\]
The constraints are given as:
\begin{itemize}
\item (C1) $0\leq i\leq1$ 
\item (C2) $0\leq u\leq1$ 
\item (C3) $0\leq x\leq i$ 
\item (C4) $0\leq y\leq1-i$ 
\item (C5) $0\leq s\leq x+u$ 
\item (C6) $max\{0,s-u\}\leq w\leq min\{s,x\}$ 
\item (C7) $max\{0,s-u\}\leq w'\leq min\{s,x\}$ 
\item (C8) $0\leq t\leq1-u+y$ 
\item (C9) $0\leq I\leq1$ 
\item (C10) $0\leq U\leq1$ 
\item (C11) $0\leq X\leq I$ 
\item (C12) $0\leq Y\leq1-I$
\item (C13) $0\leq S\leq X+U$
\item (C14) $max\{0,S-U\}\leq W\leq min\{S,X\}$ 
\item (C15) $max\{0,S-U\}\leq W'\leq min\{S,X\}$ 
\item (C16) $max\{0,t-(1-u)\}\leq o\leq min\{t,y\}$ 
\item (C17) $0\leq T\leq1-U+Y$ 
\item (C18) $max\{0,T-(1-U)\}\leq O\leq min\{T,Y\}$ 
\item (C19) $0\leq d\leq2$
\item (C20) $max\{0,T-(1-U)\}\leq O'\leq min\{T,Y\}$ 
\item (C21) $0\leq a\leq d$
\item (C22) $max\{0,d-2+u\}\leq l'\leq min\{d,u\}$ 
\item (C23) $max\{0,t-(1-u)\}\leq o'\leq min\{t,y\}$ 
\item (C24) $max\{0,d-2-l'+u+U\}\leq L'\leq min\{d-l',U\}$ 
\item (C25) $0\leq A\leq2-d$
\item (C26) $max\{0,d-2+u\}\leq l\leq min\{d,u\}$ 
\item (C27) $max\{0,d-2-l+u+U\}\leq L\leq min\{d-l,U\}$ 
\item (C28) $max\{0,a-d+l+L\}\leq j\leq min\{a,l+L\}$
\item (C29) $max\{0,d-l-L-1+U\}\leq q\leq min\{d-l-L,1-u\}$
\item (C30) $max\{0,a-d+l'+L'\}\leq j'\leq min\{a,l'+L'\}$
\item (C31) $max\{0,2-u-d+l'+L'-U\}\leq J'\leq min\{A,u-l'+U-L'\}$ 
\item (C32) $max\{0,d-l'-L'-1+U\}\leq q'\leq min\{d-l-L',1-u\}$
\item (C33) $max\{0,2-u-d+l'+L'-U\}\leq J\leq min\{A,u-l'+U-L'\}$ 
\end{itemize}
\section{}
In Eq. \eqref{eq21} the combinatorial factor $\tilde{B^{x}}_{nc}(z)$ is given as:
\begin{equation*}
    \begin{split}
        \tilde{B_{nc}^{x}}=&\underset{k\epsilon\kappa_{nc}}{\Pi}(-1)^{-w-o-w'-o'-W-O-W'-O'+l'+q'+j+l+q}\\
       & \times (-1)^{+J+j'+J'}\begin{pmatrix}a_{k}\\
b_{k}
\end{pmatrix}
    \end{split}
\end{equation*}

here, 
\begin{equation*}
    \begin{split}
        \kappa_{nc}=&\{(1,i),(1,i),(1,u),(1,u),(i,x),(1-i,y),
        (i,x),(1-i,y),\\
        &\times (x,w),(y,o),
(u,s-w),(1-u,t-o),(x,w\prime),(y,o\prime),\\
&\times (u,s-w\prime),(1-u,t-o\prime),(1,I),
(1,I),(1,U),(1,U),\\
&\times (I,X),(1-I,Y),(I,X),(1-I,Y),(X,W),(Y,O),\\
&\times (U,S-W),(1-U,T-O),(X,W'),(U,S-W'),(u,l),\\
&\times(1-u,q),(1-U,T-O')(1-U,d-l-L-q),(U,L),\\
&\times(Y,O'),(U,L'),(u,l'),(1-U,d-l'-L'-q'),\\
&\times(1-u,q'),(l+L,j),(u-l+U-L,J),(l'+L',j'),\\
&\times(2-u-d'+l'+L'-U,A'-J'),(d-l-L,a-j),\\
&\times(d-l'-L',a-j'),(2-u-d+l+L-U,A-J),\\
&\times(u-l'+U-L',J')
    \end{split}
\end{equation*}

The physical system parameters and the complex phase interference:
\begin{equation*}
    \begin{split}
        \mathcal{G}_{nc}(r,\eta_{CH},\eta,\eta')=&\frac{r^{2(i+u+I+U)}}{(1+r^{2})^{4}}\sqrt{\eta_{CH}}^{2(x+y+X+Y)}(1-\eta')^{2-a-A}\\
        &\times \frac{1}{\sqrt{12}}^{12+2x+2y+2X+2Y}(1-\eta)^{2+s+t-i+S+T-I}\\
        &\times \sqrt{1-\eta_{CH}}^{4-2(x+y+X+Y)}
    \end{split}
\end{equation*}

The detection response function is given as:

\[
\mathcal{N}_{nc,x}=f_{x+u-s}^{nc,x}f_{X+U-S}^{nc,x}f_{1+y-t-u}^{nc,x}f_{1+Y-T-U}^{nc,x}f_{i}^{nc,x}f_{I}^{nc,x}f_{a}^{nc,x}f_{A}^{nc,x}
\]

where,

\[
f_{i}^{nc,x}=i\eta(1-\eta)^{i-1}+p_{d}(1-\eta)^{i}
\]

and,
\[
f'{}_{i}^{nc,x}=i\eta'(1-\eta')^{i-1}+p_{d}(1-\eta')^{i}
\]
The parameter $z$ is used to represent the summation space $\Omega_{nc,x}$ and it's given as:
\begin{equation*}
    \begin{split}
        z=&(i,u,x,y,s,w,w\prime,t,o,o',I,U,X,Y,S,W,W',T,d,d',O,O',l,a,a',\\
        &A,A',l',L',L,q'q,j,j',J,J')
    \end{split}
\end{equation*}

and,

\[
\Omega_{nc,x}=\{z\epsilon Z_{\geq0}^{36}:\mathrm{constraints(C1)-(C36)}\}
\]

\begin{itemize}
\item (C1) $0\leq i\leq1$
\item (C2) $0\leq u\leq1$
\item (C3) $0\leq x\leq i$
\item (C4) $0\leq y\leq1-i$
\item (C5) $0\leq s\leq x+u$
\item (C6) $max\{0,s-u\}\leq w\leq min\{s,x\}$
\item (C7) $max\{0,s-u\}\leq w'\leq min\{s,x\}$
\item (C8) $0\leq t\leq1-u+y$
\item (C9) $max\{0,t-(1-u)\}\leq o\leq min\{t,y\}$
\item (C10) $max\{0,t-(1-u)\}\leq o'\leq min\{t,y\}$
\item (C11) $0\leq I\leq1$
\item (C12) $0\leq U\leq1$
\item (C13) $0\leq X\leq I$
\item (C14) $0\leq Y\leq1-I$
\item (C15) $0\leq S\leq X+U$
\item (C16) $max\{0,S-U\}\leq W\leq min\{S,X\}$
\item (C17) $max\{0,S-U\}\leq W'\leq min\{S,X\}$
\item (C18) $0\leq T\leq1-U+Y$
\item (C19) $0\leq d\leq2$
\item (C20) $0\leq d'\leq2$
\item (C21) $max\{0,T-(1-U)\}\leq O\leq min\{T,Y\}$
\item (C22) $max\{0,T-(1-U)\}\leq O'\leq min\{T,Y\}$
\item (C23) $max\{0,d-2+u\}\leq l\leq min\{d,u\}$
\item (C24) $0\leq a\leq d$
\item (C25) $0\leq a'\leq d'$
\item (C26) $0\leq A\leq2-d$
\item (C27) $0\leq A'\leq2-d$'
\item (C28) $max\{0,d'-2+u\}\leq l'\leq min\{d',u\}$
\item (C29) $max\{0,d'-2-l'+u+U\}\leq L'\leq min\{d'-l',U\}$
\item (C30) $max\{0,d-2-l+u+U\}\leq L\leq min\{d-l,U\}$
\item (C31) $max\{0,d'-l'-L'-1+U\}\leq q'\leq min\{d'-l'-L',1-u\}$
\item (C32) $max\{0,d-l-L-1+U\}\leq q\leq min\{d-l-L,1-u\}$
\item (C33) $max\{0,a-d+l+L\}\leq j\leq min\{a,l+L\}$
\item (C34) $max\{0,a'-d+l'+L'\}\leq j'\leq min\{a',l'+L'\}$
\item (C35) $max\{0,2-u-d'+l'+L'-U\}\leq J'\leq min\{A',u-l'+U-L'\}$
\item (C36) $max\{0,2-u-d+l+L-U\}\leq J\leq min\{A,u-l+U-L\}$
\end{itemize}
\end{appendices}
\end{document}